\documentstyle{l-aa}

\begin{document}

\thesaurus {
             03
            (11.09.1 NGC 3808B;
             11.09.1 NGC 6286;
             11.09.2;
             11.16.1)
}

\title{ Photometric study of polar-ring galaxies.}

\subtitle{ III. Forming rings}

\author{V.P.~Reshetnikov\inst{1}, V.A.~Hagen-Thorn\inst{2}, 
V.A.~Yakovleva\inst{2} }

\offprints{V.P.~Reshetnikov }

\institute{
       Astronomical Institute of St.Petersburg State University,
       198904 St.Petersburg, Russia
\and
       St.Petersburg State University, 198904 St.Petersburg, Russia}

\date{Received 16 January 1996; accepted 29 March 1996}

\maketitle

\begin{abstract}
We present the results of detailed surface photometry of NGC 3808B and
NGC 6286 - two spiral galaxies with possibly forming ring-like
structures rotating around major axes of the galaxies. 
The formation of rings in NGC 3808B and NGC 6286 being accompanied by 
accretion
of matter on galactic disk results in some interesting
gasdynamical and stellardynamical effects in these galaxies.
One can note, for instance, peculiar rotation curve of NGC 3808B
gaseous disk; strong infrared and H$\alpha$ emission from the galaxies;
bending and flaring stellar disks in both galaxies. Our observations
clearly illustrate the possibility that polar-ring galaxies may be
formed as a result of matter accretion from one galaxy to another.

\keywords{ galaxies: interactions; photometry; NGC 3808B; NGC 6286 }

\end{abstract}

\section{Introduction}

The general picture of the observed part of the Universe has been
in a marked degree due to the gravitational interactions and
mergings between galaxies. Observations and model calculations
show that mutual interactions and accretion may change such
fundamental galaxy characteristics as mass and luminosity
distributions, morphological type, the rate of star formation, etc.

Galaxies with polar rings (PRGs) keep aloof among the objects
which traditionally are considered as a result of interaction
between galaxies. Indeed, close encounters are the reason
of disturbance of the galaxy form and appearance of asymmetric
structures such as bridges, tails and so on, while the merging
between the galaxies is considered to result in the origin of
smoothed elliptical-like objects where the remnants of
merged objects are well mixed. To the contrary, "classic" PRGs
(category A in Whitmore et al. 1990 (PRC)) are symmetric
objects in which the remnants of interacting galaxies are not
mixed but stay in a quasi-stationary state for a
long time, may be comparable with Hubble time. Detailed
investigation of PRGs supply the new approaches to the study
of such important problems as the shape of galactic potential,
the frequency of interactions and mergings, influence of
interactions on global structure of galaxies, on their nuclear
activity etc.

Probably the most important problem concerning PRGs is their
origin. It is usually contended that the collapse of single
protogalactic cloud cannot create the object with two nearly
orthogonal large-scale kinematic systems and therefore in
the history of such objects some second event (for instance,
interaction or merging) has occured (Schweizer et al. 1983,
Whitmore et al. 1987). From the other side, recent simulations
by Curir \& Diaferio (1994) seem to indicate that a single
dissipationless collapse of rotating triaxial system can
produce misaligned spins.

Here we give the results of new observations of two candidates to
PRGs from the PRC which obviously demonstrate the possibility
of polar-ring galaxy formation as a result of interaction between 
galaxies.

Throught this paper the value $\rm H_{0}\,=\,$75 km/s/Mpc is
adopted.

\section{Observations and reductions}

The photometric observations of galaxies were carried out
in 1992 and 1994 in the prime focus of the 6-m telescope
of the Special Astrophysical Observatory of Russian Academy 
of Sciences. The photon collector was virtual phase CCD
of 520$\times$580 pixels, each 18$\times$24
$\mu$m (0.15$\times$0.2$\arcsec$) (Borisenko et al.1991).
Reimaged optics, that realized a field of view $\rm 4.5\,\times\,6.9'$
with angular resolution $\rm 0.53\,\times\,0.71"$, was used 
during our observations of NGC 6286. Observations of NGC 3808
were obtained without reimaged optics $-$ with a field of 
view $\rm 1.3\,\times\,2.0'$ and resolution $\rm 0.15\,\times\,
0.2"$. The data were acquired with standard Johnson $B$, $V$ and
Cousins $R$ filters for NGC 6286, and with Johnson $V$ and
Cousins $R$, $I$ filters for NGC 3808. Photometric calibration
was accomplished using repeated observations of the $BVRI$
standard stars from Landolt (1983) and Smith et al. (1991).
The seeing during our observations was about 2$\arcsec$. 
A log of observations is given in
Table 1 (extinction corrected sky brightnesses in each frame
in $\rm mag\,arcsec^{-2}$ are presented in the
last column of the table).
Reduction of the raw CCD data was made in standard manner
(for details see Paper I -
Reshetnikov et al. 1994). The ESO-MIDAS package
(MIDAS User Guide 1994) was used.

\begin{table}
\caption[1]{Observations} 
\begin{tabular}{llllll}
\\ \\
\hline
\\
Object&Data&Band-& Exp. & Air- & Sky \\
      &    &pass & (sec)& mass & mag. \\ \\
\hline \\
NGC 3808    & 10/11.03.94 & $V$ & 600 & 1.075 & 21.14   \\
            &             & $R$ & 300 & 1.079 & 20.66   \\
            &             & $I$ & 600 & 1.075 & 19.24   \\ \\
NGC 6286    & 6/7.05.92   & $B$ & 1000& 1.049 & 21.94 \\
            &             & $V$ & 600 & 1.043 & 21.19 \\
            &             & $R$ & 300 & 1.040 & 20.58 \\ \\
\hline
\end{tabular}
\end{table}

Spectral observations of the galaxies were obtained earlier at
the 1.93-m telescope of the Observatoire de Haute-Provence, using
the spectrograph CARELEC and the CCD TK512. For details of
observations and reductions see Reshetnikov \& Combes (1994) (RC).

\section{Results and discussion}

\subsection{NGC 3808B (PRC D-19)}

NGC 3808 (Arp 87, VV 300) is M~51-type interacting double system,
which consists of Sc spiral with $B_{t}\rm\,=\,14.1\pm0.3$
(de Vaucouleurs et al. 1991, RC3) seen nearly
face-on and a companion located at the end of spiral arm of the main
galaxy and seen edge-on. In what follows we shall refer
to the main galaxy as NGC 3808A and the companion as NGC 3808B.
The distance to the system is $\rm D\,=\,$93 Mpc (RC) and the angular
separation between companions is about 1.1' (30 kpc). Our unpublished
observations with 6-m telescope show that a radial velocity difference
between two galaxies is only 33 km/s. 

As one can see in Fig.1 where a contrast
copy from the "Atlas of Peculiar Galaxies" (Arp 1966) is given, the
spiral arm of the main galaxy is not only drawn out to the companion
but twines it forming an unrolled spiral. For this reason 
Schweizer (1986) has supposed that in the case of NGC 3808B we observe
the formation of polar ring as a result of mass transfer from the
nearby galaxy. This suggestion is confirmed by spectral observations
which have shown that the main body of the galaxy rotates around minor
axis whereas the matter of twined arm moves around the major axis
of the galaxy as in classic PRGs (Reshetnikov \& Yakovleva 1990).

\begin{figure}
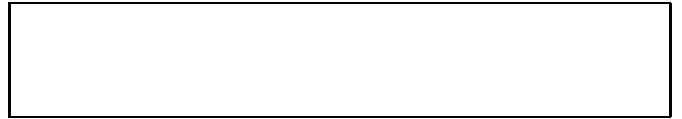

\picplace{1.5cm}
\caption[1]{Reproduction of NGC3808 from the Atlas of Peculiar Galaxies by
Arp (1966). North is at the top, east to the left}
\end{figure}

\subsubsection{Photometric characteristics}

Global photometric parameters of NGC 3808B are summarized in Table 2
(the value of $B_{t}$ in the table was derived from our $V_{t}$
magnitude and colour index from Gavazzi et al. (1991):
$B-V\,=\,$+0.65).

\begin{table*}
\caption[2]{General characteristics}
\begin{tabular}{lll}
\\ \\
\hline
\\
             & NGC 3808B &  NGC 6286 \\ \\
\hline \\
D (Mpc) [RC] & 93.0      &  76.5        \\
$A_{g}(B)$ [RC3] &  0.00     &   0.02    \\
$B_{t}\,\, (\pm\,0.10)$& 15.04  &  14.2   \\
$(B-V)_{t}$  & $\rm +0.65^{[1]}$   & $\rm +0.75\,\pm\,0.05$   \\
$(V-R)_{t}$& $\rm +0.48\,\pm\,0.03$& $\rm +0.50\,\pm\,0.03$  \\
$(R-I)_{t}$& $\rm +0.46\,\pm\,0.03$ \\
$ a\,\,(\mu_{B}=\rm 26)$ & 0.77' (20.7 kpc) & 1.68' (37.5 kpc) \\
$ b/a\,\,(\mu_{B}=\rm 26$)& 0.36:   & 0.36:      \\ \\
$\mu_{0}$  &  18.5($V$)  &  20.4 ($ B)^{\rm [2]}$  \\
$h$           & 3.9" (1.8 kpc)&  6.6" $\rm (2.5\,\,kpc)^{[2]}$ \\
$z_{0}$ & 2.2" $\rm (1.0\,\,kpc)^{[3]}$& 3.8" $\rm (1.4\,\,kpc)^{[3]}$ \\
$\rm V_{max}$ (km/s) & 120$\pm$10        \\ \\
$Possible\,\,forming\,\,ring:$ \\
$\rm M $ &  -18($V$)  &     -18.4($B$)      \\
$ B-V $ &           &   +0.5:               \\
$ V-R $ &  +0.4:  &      +0.3:               \\
$ R-I $ &  +0.4:  & \\
$\rm V_{max} (km/s) $& 120  & 100:\\
Ring-to-central galaxy\\
ratio                    &   0.1($V$) &  0.2($B$)         \\ \\
\hline
\end{tabular}

(1) Gavazzi et al. (1991)  \\
(2) Value determined from the equivalent profile        \\
(3) Minimum value observed within galactic disk           \\
\end{table*}

There are some published results of photometric observations of NGC 3808B.
Tomov (1978) has found $V\,=\,$14.27 for the 54" aperture which covers
almost whole galaxy. This value is more bright than that found by us
($V_{t}\,=\,$14.39) probably because of getting in the aperture the
star, located to SW from the galaxy.

Gavazzi et al. (1991) give the results of photoelectric observations
of NGC 3808B with three various apertures. The mean difference
between these values and our measurements for the same apertures
is $\rm -0.05\pm0.02\,(\sigma)$. In the work of Gavazzi \& Randone (1994)
the apparent magnitude in the $V$ band within the elliptical
isophote of 25 $\rm mag\,arcsec^{-2}$ is given: $V_{25}\,=\,$14.61. Taking
in mind the mean difference between their $V_{25}$ and total magnitudes
in RC3 ($\rm +0.3\pm0.2$) we make the conclusion about good agreement 
between our $V_{t}$ value and the data of Gavazzi \& Randone (1994).

The isophotes of NGC 3808B in the $V$ band are given in Fig.2. The main
body of the galaxy is viewed practically edge-on. The twined arm become 
apparent
as "swelling" of isophotes to NW and SE from the main body. Outside
the main body the arm does not end but goes to NW from its edge
as far as $\approx$40" (18 kpc). In Fig.2 one can see faint ($V\,=\,$18.7)
galaxy projected on the NE edge of NGC 3808B. We think this galaxy
is background one seen here by chance. However, it may be
gravitationally bound with NGC 3808A,B but because of its non-distorted
structure it is located sufficiently far from the pair and does not
participate in strong gravitational interaction between A and B galaxies.

\begin{figure}
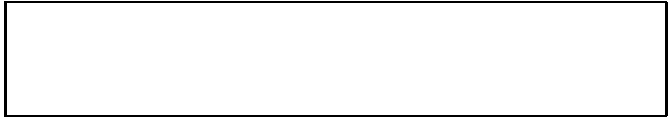

\picplace{1.5cm}
\caption[2]{Isophotal contour image of NGC 3808B in the $V$ passband.
The faintest contour is 26.0 $\rm mag\,arcsec^{-2}$, isophotes step -
$\rm 0.^{m}75$. The arrow indicates a length of 60"}
\end{figure}

\subsubsection{Host galaxy}

Observed colour indexes of NGC 3808B listed in Table 2 correspond to the
normal Sab-Sb galaxy. After reducing for redshift and inclination made
in accordance with RC3 recommendations the colour indexes ($B-V\approx$+0.5,
$V-R\approx$+0.4) correspond to Sc-Scd galaxy (Buta et al. 1994, 
Buta \& Williams 1995). 

The photometric profile along the major axis ($\rm P.A.=55^{o}$)
is shown in Fig.3a. As one can see, the surface brightness distribution 
is nearly exponential (note the contribution  
of star and faint galaxy in SW and NE parts of the
profile correspondingly at $\rm r\geq20"$). Representing the profile
by exponential law we find that the parameters of the brightness
distribution for $\rm r\le20"$ are $h\,=\rm\,3.9"\pm0.35"$
(exponential scale-length) and $\mu_{0}(V)\,=\rm\,18.5\pm0.2$ (central
surface brightness). After transforming $\mu_{0}(V)$ value to 
$\mu_{0}(R)$ (in the Cousins system) by assuming that $V-R\,=\,$0.5 we 
have compared the disk parameters of NGC 3808B with those for large
sample of interacting galaxies studied in Reshetnikov et al. (1993b).
The photometric parameters of NGC 3808B are found to be typical for
extremly interacting galaxies (Figs.13 and 15 in cited paper).

\begin{figure}
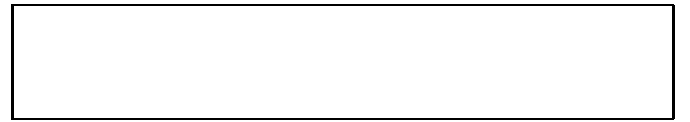

\picplace{1.5cm}
\caption[3]{(a) Luminosity profile along the major axis of NGC 3808B
in the $V$ passband. (b) Distribution of radial velocities along
the major axis of the galaxy}
\end{figure}

\begin{figure}
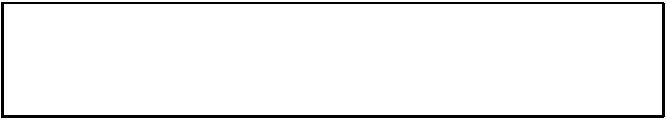

\picplace{1.5cm}
\caption[4]{(a) Photometric profile of NGC 3808B along the minor
axis. (b) Radial velocities distribution along the minor axis
of the galaxy}
\end{figure}

It is well known that in normal spirals the z-structure of stellar
disks is well represented by the model of a locally isothermal,
self-gravitating stellar sheet with scale-height $z_{0}$ independent
on galactocentric distance (van der Kruit \& Searle 1981). The disks
of non-dwarf spiral galaxies can be characterized by the value of
$z_{0}\,=\rm\,0.7\pm0.2$ kpc (van der Kruit 1989). The distributions
of $z_{0}$ values along the disk of NGC 3808B in three passbands are
given in Fig.5. To eliminate the influence of twined arm and surrounding
envelope, only the most bright inner parts of profiles parallel to the
minor axis were used for estimating the scale-height values of the disk. 
As one can see, these values increase with moving away from the center
of galaxy (at the edge of the disk $z_{0}$  is approximately twice as
large as at the center). Such stellar disk structure is expected
for the galaxies subjected to strong external accretion or merging
(Toth \& Ostriker 1992, Quinn et al. 1993).

\begin{figure}
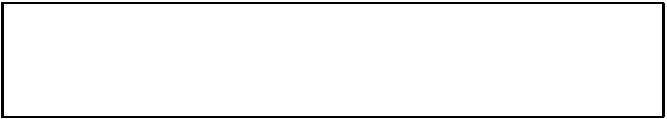

\picplace{1.5cm}
\caption[5]{Scale-height distributions for NGC 3808B as a function of
radius along the major axis. Circles represent the data in the $V$ band,
crosses - $R$, rectangles - $I$}
\end{figure}

The rotation curve of NGC 3808B along
the major axis constructed using the data of emission lines H$\alpha$
and [N~II] is shown in Fig.3b according to RC.
It is seen that the galaxy demonstrates very peculiar
kinematics: the kinematic center of the galaxy is shifted from the
photomteric one by approximately 0.7" (0.3 kpc) to NE; after reaching the
maximum value ($\rm V_{max}\,=\,120\pm10$ km/s) at $\rm r\approx3"$ the
radial velocities show rapid decrease to the value nearly equal to the
systematic velocity. According to classification of rotation curves for
interacting galaxies suggested by Keel (1993) (Fig.2 in that paper) the
rotation curve of NGC 3808B belongs to the group DIST (disturbed).

In Fig.6 the solid line represents the rotation curve of the galaxy
averaged relative to the center and polinomially smoothed. Evidently
it cannot reflect a real mass distribution in the galaxy but is
a result of combined influence of gravitational
disturbance from massive companion (NGC 3808A) and matter accretion
from that to NGC 3808B on the gaseous disk of the latter.
Let us find what rotation curve NGC 3808B would have
if it was not a member of interacting pair. To do this we
have approximated the inner part ($\rm r\leq2.5"$) of the observed
rotation curve by the model of exponential disk (Monnet \& Simien 1977)
with intrinsic axial 
ratio $b/a\,=\rm\,0.2$ and the value of scale-length $h\,=\rm\,3.9"$ 
found earlier. The resulting curve depending
besides $b/a$ on the only free parameter (mass-to-light ratio which is
$f_{V}\,=\,$3.6 in solar units) is shown in Fig.6 by dashed line. 
This curve is hypothetical "non-disturbed" rotation curve of the galaxy.
The maximum velocity of the model rotation curve is equal to 173 km/s
at $\rm r=9"$ (4.1 kpc). Let us consider how it agrees with other 
global characteristics of the galaxy. 

\begin{figure}
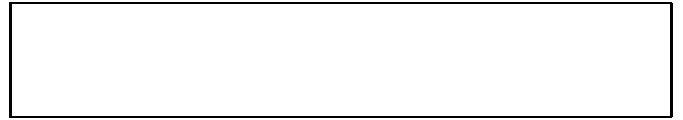

\picplace{1.5cm}
\caption[6]{Averaged rotation curve of the main body of NGC 3808B 
(solid line) and of the forming ring (short-dashed line). 
Dashed line represents model of "non-disturbed" rotation curve}
\end{figure}

As it follows from integral colour indexes and surface brightness
distribution, NGC 3808B belongs to Sc-Scd galaxies. Using the
data on 207 spiral galaxies compiled by Corradi \& Capaccipli (1991),
Reshetnikov (1994) found the following dependence between numerical
Hubble stage index ($T$) and maximum rotation velocity:
$\rm V_{max}\,(km/s)\,=\,292\,-\,22.7$$T$. From this equation the
value $\rm V_{max}\,\approx\,156-179$ km/s should be expected for
NGC 3808B. The value of $\rm V_{max}$ for "non-disturbed" rotation
curve falls into this interval. The Tully-Fisher relationship in the
$B$ band according to Rubin et al. (1985) shows that normal Sc galaxy
with the luminosity of NGC 3808B must have the maximum rotational
velocity of 160$-$165 km/s which is also in agreement with our model 
rotation curve. The use of the Tully-Fisher relationship in the $I$
band according to Byun (1992) leads to better agreement - 
$\rm V_{max}\,\approx\,170$ km/s.

Comparing observed and "non-disturbed" rotation curves one can conclude
that strong gravitational disturbance together with mass transfer
to NGC 3808B results in observed displacement of the global maximum
of rotation curve to the center (approximately by $\rm 6"=2.7$ kpc) and
in "braking" (approximately by 50 km/s) of the gaseous disk of the
galaxy. The latter is in agreement with the conclusion of Reshetnikov (1994)
who found that spiral galaxies in strongly interacting systems tend to
have lower observational rotational velocities than the same field galaxies.
 
Interaction between gaseous disk of the galaxy and accreting gas of
twined spiral arm may lead (besides of disturbance of the velocity
field) to intensification of star-forming activity in NGC 3808B.
Indeed, as it was pointed out in RC, NGC 3808B shows both strong
emission lines in circumnuclear regions (within rectangular aperture
$\rm 2.5"\times2.3"$, centered on the nucleus, the equivalent width
of H$\alpha$ emission line and the observed H$\alpha$ luminosity
are found to be W(H$\alpha)\,=\,39\,\rm\AA$,
$L$(H$\alpha)\,=\,2\,\times\,10^{40}$ erg/s)
and intense far-infrared radiation ($L\rm (FIR)\,=\,4.4\,\times\,
10^{10}$$L_{\sun}$). For a sample of interacting galaxies Keel (1993)
has found a correlation between H$\alpha$ equivalent width and greatest
observed deviation from a symmetric rotation curve when normalized to
the peak rotation velocity. Comparing the data for NGC 3808B with Fig.5
in Keel (1993) we find that the parameters of the galaxy are in agreement
with the dependence presented in that paper.

To consider the problem of emission mechanism in the nucleus of NGC 3808B
we compiled the results of spectral observations of the galaxy from
Keel et al. (1985) and RC. The reddening was taken into account by
standard way adopting $I$(H$\alpha$)/$I$(H$\beta$)$\,=\,$3.1 (Veilleux
\& Osterbrock 1987). The value of absorption in the direction of the 
nucleus was found to be $\rm 3.^{m}4$ in the $B$ band. It turned out
that in standard classification
diagrams ( [O~III]$\lambda$5007/H$\beta$ vs. [N~II]$\lambda$6583/H$\alpha$
and [O~I]$\lambda$6300/H$\alpha$, one of these is given in Fig.7) 
reddening-corrected line ratios are located near
the boundary between AGNs (Seyferts and LINERs) and HII-like
objects (Veilleux \& Osterbrock 1987). We think that the nucleus of
NGC 3808B may be an example of masking of active nucleus by circumnuclear
burst of star formation - the situation discussed earlier by 
Keel et al. (1985).
(Note that the reddening-corrected H$\alpha$ luminosity of the central
region of the galaxy is more than $\rm 10^{41}$ erg/s.) For the final
decision of the problem of the nature of NGC 3808B nucleus spectral
observations with high (better than 1") spatial resolution are needed.

\begin{figure}
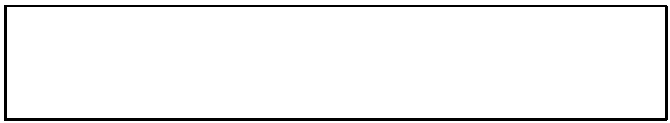

\picplace{1.5cm}
\caption[7]{Reddening-corrected 
[OIII]$\lambda$5007/H$\beta$ vs. [OI]$\lambda$6300/H$\alpha$
intensity ratio
for the nucleus of NGC 3808B (circle). Solid curve divides AGNs from
HII region-like objects according to Veilleux \& Osterbrock (1987)}
\end{figure}

\subsubsection{Forming ring}

The photometric profile of NGC 3808B along the minor axis is shown in
Fig.4a. Just for $\rm r\approx5"$ from the nucleus surface 
brightness distribution becomes more flat than that usually observed in
normal edge-on spirals. This feature may be explaned by the contribution
to the observed profile of the most bright and contrast part of spiral
arm of NGC 3808A drawn out to NGC 3808B and twined the latter (see Fig.1).
Surface brightness of this pseudo-ring structure crossing the central
region of NGC 3808B along its minor axis is $\mu(V)\,\approx\rm\,23-24$.

The dependence on distance from the nucleus of radial velocities along 
the minor axis found in RC from measurements of H$\alpha$ and [N~II]
lines is given in Fig.4b. As one can see, the rotation curve of the
forming ring represents almost a straight line (see also Fig.6). The slope
of the rotation curve is equal to 13 km/s/arcsec or 29 km/s/kpc, 
the maximum velocity is close to that observed in the main galaxy (Fig.6)
(such a feature is typical for "classic" PRGs).
The straight character probably suggests that a relatively narrow ring
has been formed in circumpolar plane of the galaxy. 
The possibility of forming such ring structures accumulating
the part of accreting matter has been shown by Weil \& Hernquist (1993)
and Sotnikova (1996)
in course of numerical simulations. The gas of the ring
interacting with that of the galaxy disk settles on it. On the other
hand there is a permanent supply of the ring with the matter transfered 
from NGC 3808A. Stopping of accretion seems to result in prompt 
disappearence of the ring.  

The maximum observed rotational velocity in the forming ring (as well as
in the main body of the galaxy) is approximately equal to 120 km/s.
Its total (after account for all parts twined around NGC 3808B) observed
absolute luminosity is $M_{V}\,\approx\rm\,-18$ that is about 10\% of
total galaxy luminosity.  The observed colour indexes of unrolled spiral arm
($V-R\,=\,$+0.4, $R-I\,=\,$+0.4) are more blue than colour indexes of the
main galaxy. The mean colours of the pseudo-ring indicate an age of 
about $\rm 10^{9}$ years, for a solar metallicity. The matter of the
pseudo-ring probably contains some dust since in the region of the spiral arm
projection to the SW edge of main body the local
increase of colour indexes (with $E_{V-R}\,=\rm\,0.^{m}1-0.^{m}15$) is
observed. In addition, visual inspection of the reproduction of this system in
the Atlas by Arp (1966) shows that there is some absorption in this region.

The whole projected length of twined arm ($\geq$2'$=$55 kpc) gives a 
possibility
to estimate the duration of twisting of accreted matter around NGC 3808B. 
Adopting that a velocity of the gas is equal to maximum observed velocity
in the forming ring (120 km/s), one can find that the duration of active
phase of interaction with mass transfer is at least $\rm 4\times10^{8}$
years. Taking into account the projection factor this period may be 
increased up to $\rm 10^{9}$ years. This value corresponds approximately
to the stellar population age in the forming ring. Therefore, in this
case we deal with gravitationally bound galaxies interacting over a long
time but not with chance encounter. In many aspects NGC 3808
resembles interacting system NGC 7753/52 (Arp 86) in which mass
transfer occurs from main galaxy to low-massive companion (Laurikainen 
et al. 1993, Salo \& Laurikainen 1993). As in NGC 3808B, the accretion
of matter to NGC 7752 results in global bursts of star formation in this
galaxy. Numerical modeling shows that the components of Arp 86 are
moving in a low-eccentricity relative orbit. 
By analogy, we suggest that the members of NGC 3808
are also at nearly circular relative orbit favourable to mass transfer.

\subsection{NGC 6286 (PRC C-51)}

NGC 6285/6 (Arp 293) is an interacting double system resembling
NGC 3808A,B. The mean difference between them is that in the
case of NGC 6285/6 both galaxies have comparable masses.
For $\rm D\,=\,$76.5 Mpc (RC) the angular distance between galaxies
(1.5') corresponds to the projected distance of 33 kpc.
The radial
velocity difference is not well known but seems to be no more
than 200 km/s (NED\footnote{The NASA/IPAC Extragalactic Database (NED)
is operated by the Jet Propulsion Laboratory, California Institute of
Technology, under contract with the National Aeronautics and Space
Administration.}). 

In Fig.8 the reproductions of $B$-band
CCD image of NGC 6286 (just this galaxy is a candidate to PRG)
are shown in a logarithmic scale with different contrasts for
better displaying optical structure of the galaxy.
The main body of NGC 6286 is seen to be a spiral galaxy viewed
practically edge-on and crossed by bended dust strip (Fig.8a).
In Fig.8b a diffuse semi-ring extending from SE edge of the
galaxy is well notable (see also reproductions in Arp (1966) and
in the PRC). In Fig.8c one can see also a weak bridge between
two galaxies. The second galaxy of the pair (NGC 6285) is an
early-type spiral ($T=$-1 according to RC3) with
$B_{t}\,=\rm\,14.48\pm0.15$ (RC3). 

\begin{figure}
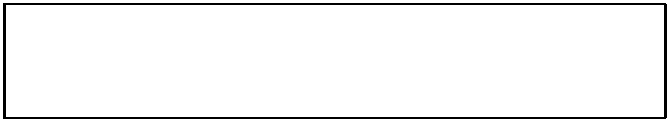

\picplace{1.5cm}
\caption[8]{$B$-band image of NGC 6286 displayed at three different
contrasts (in a logarithmic gray scale). The entire image is 
1.85$\times$2.1 arcmin}
\end{figure}

In RC the rotation curve of emitting gas along the minor axis
of NGC 6286 is given. Though curve looks very peculiar
it demonstrates possible large-scale
rotation of the matter around the major axis of the galaxy. Therefore,
NGC 6286 may be related (as NGC 3808B) to galaxies with forming
polar rings.

\subsubsection{Photometric characteristics}

The main photometric parameters of NGC 6286 are collected in
Table 2. 

The total magnitude of the galaxy according to RC3 is
$B_{t}\,=\rm\,14.06\pm0.18$. This value is somewhat brighter
than found here ($B_{t}\,=\rm\,14.2\pm0.1$) but within the
quoted errors this difference is insignificant. The total
magnitude of NGC 6286 in $R$ band ($R_{t}\,=\rm\,12.95$) found
in this work is in good agreement with that given by
Reshetnikov et al. (1993a) ($R_{t}\,=\rm\,12.89$).

The isophotes of NGC 6286 in the $V$ passband are shown in
Fig.9. As in the case of NGC 3808B, the main body of the
galaxy is viewed nearly edge-on. The abovementioned dust strip
results in strong isophotes distortion of the central part
of the galaxy. The ring structure forming around the galaxy
is well turned to the line of sight. The major axis of this
possible ring is about 1.2'$=$27 kpc by isophote $\mu_{V}\,=\rm\,
24.35$, the apparent axial ratio is $b/a\,\approx\,$0.7.

\begin{figure}
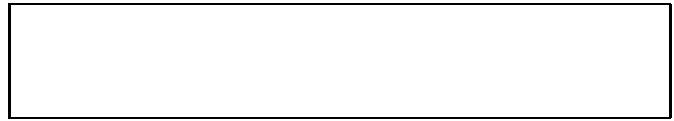

\picplace{1.5cm}
\caption[9]{
Isophotal contour image of NGC 6286 in the $V$ passband.
The faintest contour is 25.1 $\rm mag\,arcsec^{-2}$, isophotes step -
$\rm 0.^{m}75$. The arrow indicates
a length of 60".}
\end{figure}

\subsubsection{Host galaxy}

Observed colour indexes of NGC 6286 given in Table 2
correspond to the normal Sa galaxy. After standard reductions
for redshift and inclination colour indexes ($B-V\approx$+0.55 ,
$V-R\approx$+0.4) turn out to be the same
as in Sbc-Sc galaxies (Buta et al. 1994, Buta \& Williams 1995).
However, NGC 6286 is Sb galaxy according to RC3. This discrepancy
may be due to both abnormally blue colour of NGC 6286 and 
uncertainty in inclination correction for this nearly edge-on
galaxy.

In Fig.10 the photometric profile along the major axis
($\rm P.A.=33^{o}$) in the $B$ band is shown. Up to 
$\rm r\approx\pm30"$ from the nucleus the surface brightness
decreases very slowly, then falls down more rapidly. Note
that the profile shows no signs of pronounced bulge in the
central region of the galaxy (partly may be due to absorption
of the bulge emission by dust lane projected onto the central
region - see Fig.8,9). The equivalent luminosity profile of
the galaxy is shown in Fig.11. As one can see, up to
$\rm r^{*}\approx20"$ (or $\mu(B)\approx$24) the averaged 
surface brightness distribution of the galaxy is well described
by exponential law with parameters $\mu_{0}(B)\rm=20.37$,
$h^{*}=$6.6" (dashed line in Fig.11), 
then the profile changes its slope and becomes
more flat (see also the equivalent profile in the $R$ passband
published in Reshetnikov et al. (1993a)). This change at
$\rm r^{*}\approx20"$ is probably due to the contribution of
ring-like structure surrounding the galaxy. Let us note also
that the central surface brightness of the galactic disk is
considerably brighter than canonical value $\mu_{0}(B)=$21.65
(Freeman 1970). Such a property is typical for the galaxies
in interacting systems (Reshetnikov et al. 1993b).

\begin{figure}
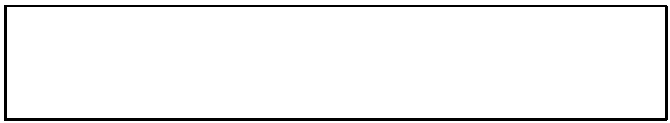

\picplace{1.5cm}
\caption[10]{Luminosity profile along the major axis of NGC 6286
in the $B$ passband}
\end{figure}

\begin{figure}
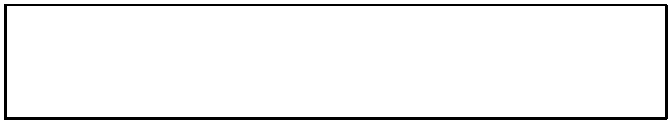

\picplace{1.5cm}
\caption[11]{Equivalent luminosity profile of NGC 6286 in the
$B$ passband}
\end{figure}

The distribution of $z_{0}$ values along the disk of NGC 6286
is shown in Fig.12 for three passbands. To exclude the 
influence of ring-like structure on the estimation of 
scale-height of the disk we use, as in the case of NGC 3808B,
only inner, most bright parts of profiles parallel to minor
axis. The central region of the galaxy disturbed by the dust
strip is also excluded. The values of scale-height are seen to
increase while shifting from the galaxy nucleus, but not so much
as in NGC 3808B: only 30\%$-$50\% for the distances from
5 kpc to 15 kpc from the nucleus.

\begin{figure}
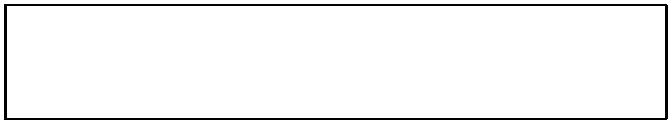

\picplace{1.5cm}
\caption[12]{
Scale-height distributions for NGC 6286 as a function of
radius along the major axis. Circles represent the data in the
$B$ passband, crosses - $V$, rectangles - $R$}
\end{figure}

As well as NGC 3808B, NGC 6286 demonstrates (see RC) both
strong emission lines in circumnuclear region (within
rectangular aperture $\rm 2.5"\times2.3"$ centered to the
nucleus W(H$\alpha)\,=\,24\,\rm\AA$, $L$(H$\alpha)\,=\,2.1\,\times\,
10^{40}$ erg/s), and strong IR emission
($L\rm (FIR)\,\approx\,10^{11}$$L_{\sun}$). This may be a
consequence of a burst of star formation induced by strong
interaction with the companion. Emission-line spectrum of
the nucleus is of HII-region type (RC).

\subsubsection{Forming ring}

Photometric profile along the minor axis of the galaxy is
shown in Fig.13a. The central part of the profile follows
to exact exponential law. For $\rm r\geq$20" the main contribution
to the observed brightness is given by the structure which we
consider to be a ring forming around the galaxy.
The ring is asymmetric: its SE part (semi-ring well seen in
Fig.8b) has mean observed surface brightness $\mu(B)\approx$24
while its NW part is fainter by approximately $\rm 1^{m}$
(as NW part we imply the nearest to NGC 6286 region of the
bridge connecting two galaxies). The total magnitude of 
forming ring in the $B$ band is $\rm 16.0\pm0.3$. Therefore,
the contribution of the ring to the total luminosity of the
galaxy is as high as 20\%. Observed integral colour indexes
of SE part of the ring ($B-V\,=\rm\,+0.56\pm0.06$, $V-R\,=\rm\,
+0.36\pm0.05$) are more blue than those of main galaxy.
But after correction for inclination of the galaxy to the
line of sight the colour indexes of both are very close.
It should be noted also that colour index $B-V$ ($\rm\approx +0.8$) 
of the bridge
between galaxies is somewhat redder than that of the main
body of NGC 6286.

The distribution of radial velocities along the minor axis of
NGC 6286 found from measurements of emission lines H$\alpha$
and [NII]$\lambda$6583 in RC (where NGC 6286 is named NGC 6285
by mistake) is shown in Fig.13b. 
One can suppose that there are two subsystems of gas in the galaxy.
One of them well seen at $\rm r\approx$20" in H$\alpha$ probably
reflects large-scale rotation of the forming ring. 
From simple geometrical arguments used
earlier for NGC 3808B, we have found that active phase of
interaction between the galaxies being accompanied by mass
transfer has been lasting no more than $\rm 5\times10^{8}$ years. 

\begin{figure}
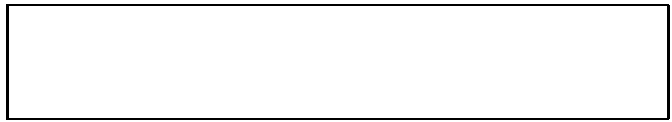

\picplace{1.5cm}
\caption[13]{
(a) Photometric profile of NGC 6286 along the minor
axis in the $B$ passband. 
(b) Radial velocities distribution along the minor axis
of the galaxy (crosses - H$\alpha$, circles - [NII]$\lambda$6583).
}
\end{figure}

\section{Conclusions}

In this work we give the results of observations of two spiral
galaxies in close interacting systems. In both systems mass
transfer from one galaxy to another is observed, accreting 
matter forming large-scale ring-like structures rotating around
the major axes of considered galaxies. Because of existence
of two kinematically distinct subsystems rotating in orthogonal
planes these objects - NGC 3808B and NGC 6286 - look very much
like polar-ring galaxies. Both galaxies clearly illustrate
the possibility that PRGs may be formed as a result of matter
accretion from one galaxy to another (Schweizer et al. 1983).

In NGC 6286 the ring structure seems to be
at relatively earlier stage of formation than in NGC 3808B.
Accreting matter in NGC 6286 probably did not have time to perform 
the whole revolution around the galaxy.

Both galaxies with forming rings are spirals and in both
cases the diameter of observed ring is comparable (or
even smaller) with diameter of the galaxy. It means that
interaction has to occur between gaseous disk of the galaxy
and the ring. Therefore, if mass transfer from the companion
stops then because of this interaction the ring structure
must fall on the main galaxy disk very soon. Let us note also
that inclined rings around spiral galaxies may be relatively
stable only in the case when their diameters are larger than
those of gaseous disks of main galaxies. ESO 235-58 (Buta
\& Crocker 1993) and NGC 660 (van Driel et al. 1995) are
possible examples of such objects. 

The formation of rings in NGC 3808B and NGC 6286 being accompanied by
accretion of matter on galactic disks and interaction of two
nearly orthogonal gaseous subsystems results in some
interesting gasdynamical and stellardynamical effects in
these galaxies. For instance, we may notice the peculiar
rotation curve of NGC 3808B gaseous disk (see Fig.3b). Both
galaxies demonstrate strong infra-red and H$\alpha$
emission pointing to the high star-forming activity.
Stellar disks of both galaxies increase their thickness
when going from the center to periphery and are bended in the
outlying regions. One can conclude that double systems
similar to those considered in this paper are unique laboratories
for studying effects of outer accretion on galaxy structure
and evolution.

In conclusion let us pay attention to the fact that both
NGC 3808B and NGC 6286 are viewed edge-on. It is such orientation
that allows to include these in PRC, based on optical
morphology. If NGC 3808B and NGC 6286 were observed at
smaller angles to the line of sight, faint diffuse ring
structures being projected onto the more bright regions of
main galaxies would be indistinguishable and then the galaxies
would not be included in PRC as candidates in PRGs. Undoubtedly,
among double interacting galaxies other systems must exist
which are similar to considered here but viewed by less
advantageous angles for discovering the ring structures. 
For the search of such objects it would be interesting
to undertake detail kinematic investigation of the sample of
M~51-type galaxies (there are 160 such systems pointed out
by Vorontsov-Velyaminov 1975).

\acknowledgements 
{We are grateful to A.Kopylov and S.Kajsin for 
assistance at the telescope
and 6-m telescope Committee for the allocation of observing time.
This research has made use of the NASA/IPAC Extragalactic
Database (NED). 
Financial support was provided in part by
grant $N$ 94-02-06026-a from the Russian Foundation for
Basic Research and grant A-03-013 from the 
ESO C$\&$EE Programme.}

\end{document}